# 3D Printing of Near-Ambient Responsive Liquid Crystal Elastomers with Enhanced Nematic Order and Pluralized Transformation

*Dongxiao Li, Yuxuan Sun\*\*, Xingjian Li, Xingxiang Li, Zhengqing Zhu, Boxi Sun, Shutong Nong, Jiyang Wu, Tingrui Pan, Weihua Li, Shiwu Zhang\*\*\*, Mujun Li#\**

# Lead contact.
\* Corresponding authors.

D.X. Li, Y.X. Sun, X.X. Li, Z.Q. Zhu, B.X. Sun, S.T. Nong, J.Y. Wu, S.W. Zhang, M.J. Li

Institute of Humanoid Robots, Department of Precision Machinery and Precision Instrumentation, University of Science and Technology of China, Hefei, 230026, China

E-mail: yuxuans@ustc.edu.cn, swzhang@ustc.edu.cn, lmn@ustc.edu.cn

X.J. Li

School of Microelectronics, University of Science and Technology of China, Hefei, 230026, China

T.R. Pan

Suzhou Institute for Advanced Research, University of Science and Technology of China, Suzhou 215123, China

W.H. Li

School of Mechanical, Materials, Mechatronic and Biomedical Engineering, University of Wollongong, Wollongong, NSW 2522, Australia

**Keywords:** 3D printing, liquid crystal elastomer, near-ambient temperature, nematic order

**Abstract:** Liquid Crystal Elastomers with near-ambient temperature-responsiveness (NAT-LCEs) have been extensively studied for building bio-compatible, low-power consumption devices and robotics. However, conventional manufacturing methods face limitations in programmability (e.g., molding) or low nematic order (e.g., DIW printing). Here, a hybrid cooling strategy is proposed for programmable 3D printing of NAT-LCEs with enhanced nematic order, intricate shape forming, and morphing capability. By integrating a low-temperature nozzle and a cooling platform into a 3D printer, the resulting temperature field synergistically facilitates mesogen alignment during extrusion and disruption-free UV cross-linking. This method achieves a nematic order 3000% higher than NAT-LCEs fabricated using traditional room temperature 3D printing.

Enabled by shifting of transition temperature during hybrid cooling printing, printed sheets spontaneously turn into 3D structures after release from the platform, exhibiting bidirectional deformation with heating and cooling. By adjusting the nozzle and plate temperatures, NAT-LCEs with graded properties can be fabricated for intricate shape morphing. A wristband system with enhanced heart rate monitoring is also developed based on 3D-printed NAT-LCE. Our method may open new possibilities for soft robotics, biomedical devices, and wearable electronics.

**Introduction**

Liquid Crystal Elastomers (LCEs), a class of smart material that combines flexible polymer networks with anisotropic liquid crystal mesogens, have garnered great attention for their programmable, reversible deformations and high energy densities.[1–4] Over the past decade, LCEs have found great promise in building soft actuators,[5–11] wearable devices[5,12,13], and robotics[14–20]. LCEs are inherently thermotropic. Different stimuli (e.g., heat, light, electricity)[14,21–26] enable LCEs to surpass their nematic-isotropic transition temperature ($T_{NI}$), disrupting the order of the aligned liquid crystal mesogens and resulting in a contraction up to 50% along their original alignment direction. LCEs typically exhibit a $T_{NI}$ ranging from 70 to 160 °C,[27–29] substantially higher than ambient temperature. Since Earth's surface temperatures usually range from -83 to 56 °C,[30–32] a high $T_{NI}$ requires more considerable input power. Since the human body's temperature has a tolerable limit of 52 °C,[33] LCEs may suffer from insufficient deformation or pose biological hazards (Figure 1A). Furthermore, this high $T_{NI}$ also hinders the possibility of actuation through environmental or body heat. Therefore, the high $T_{NI}$ of LCEs has limited their broader potential for real-world applications.

Recently, LCEs with a near-ambient transition temperature (NAT-LCEs) have been developed, expanding their use in biomedical devices and soft robotics. For example, Saed et al. synthesized LCEs with tunable $T_{NI}$ from 28 to 105 °C by controlling the ratio of liquid crystal monomer and the category of crosslinker, thereby realizing sequential and reversible shape changes.[34] Bauman et al. further reduced the $T_{NI}$ and increased the actuation rate of LCEs using a liquid crystalline diacrylate (C6BAPE) integrated into various chemical reactions.[35] Wu et al. developed LCE metamaterials with biaxial actuation strain (-53%). They reduced $T_{NI}$ to 67 °C by designing a 2D lattice pattern of straight ribbons and using mechanical prestrain followed by UV photo-polymerization to lock in the alignment.[36] Since the pre-defined mesogen alignment determines the thermally induced deformation of NAT-LCEs, programming the alignment pattern is crucial. Most NAT-LCEs have been programmed with mechanical alignment, which can be achieved by directly stretching NAT-LCEs or applying them mechanical force with a mold.[35–37] Such a method is straightforward and does not require specialized equipment. However, it can only produce NAT-LCEs with simple monodirectional or rotational alignments.[38–42] Similarly, magnetic or electric-field-induced alignment methods are theoretically possible for programming NAT-LCEs.[43,44] Still, the mesogen alignment patterns achieved with these methods are relatively simplistic, mainly due to the complexities of manipulating these fields.[45,46] Photo-induced alignment has been reported in programming LCEs with complex patterns.[47,48] Yet, its application is confined to thin films no thicker than 50 μm and necessitates a sophisticated process for generating the template surface.

In contrast, direct ink write (DIW) 3D printing is particularly attractive because of its capability of arbitrary patterning mesogen alignment within customized shapes. Before printing, an oligomer ink with suitable rheological properties is synthesized via a chain-extension reaction. During DIW, as the ink is extruded through the nozzle, the liquid crystal mesogens are subjected to strong shear stress, aligning them in the flow direction.[27,34,29,35,49] This alignment enables the 3D printing of nematic LCEs with spatially patterned director profiles determined by the print path. Consequently, this process-induced alignment allows for highly controllable and complex deformations in LCEs.[4,50] However, the 3D printing of NAT-LCEs remains a challenge. Since current DIW processes use a nozzle temperature at or above ambient temperature, the ink remains a low-viscosity liquid with low shear stress, making it difficult to align mesogens. Additionally, the alignment of mesogen requires rapid UV crosslinking after extrusion. UV crosslinking is an exothermic reaction that can result in significant thermal disruption to the ordered alignment of mesogens. These factors are interdependent and must be optimized synergistically to enhance alignment and tunability of the mesogen alignment. Consequently, a general strategy for the 3D printing of NAT-LCEs with effective mesogen alignment, arbitrary programming, and tunable alignment order is still under exploration and development.

Besides the existing challenges, an intriguing phenomenon is that the $T_{NI}$ of oligomer inks is typically about 5 to 35 °C lower than the temperature of the cured NAT-LCEs, often around or below 15 °C.[34,35] This presents a challenge for the DIW of NAT-LCEs while also introducing an exciting consideration: the change in $T_{NI}$ during the DIW process affects the shape of the as-printed NAT-LCEs. Since DIW printing of NAT-LCEs may occur within the temperature range of their shape-changing phase, the $T_{NI}$ variation could transform a printed 2D structure into a 3D one. This characteristic could potentially create sophisticated as-printed shapes and complex deformation control strategies at room or body temperature. However, it remains unexplored to date. Additionally, the transition of $T_{NI}$ endows the materials with bidirectional deformation capabilities, as liquid crystal mesogens exhibit increasing disorder at higher temperatures and enhanced order at lower temperatures. This adaptability to different thermal stimuli allows for more versatile applications.

In response to the current challenges, we present a hybrid cooling strategy that facilitates the 3D printing and programming of NAT-LCEs by manipulating their rheological and thermodynamic properties. The extruded NAT-LCE inks are kept in a nematic state with high viscosity by integrating a liquid cooler into the nozzle. A semiconductive cooler and a UV lamp (365 nm wavelength) are also utilized on the printing platform to rapidly lock the LCE alignment without disrupting the mesogen order during cross-linking. We achieved an order parameter of 0.406 for NAT-LCEs, compared to 0.013 for those printed at room temperature. A transition temperature change of approximately 30 °C was observed during the printing process, leading to the fabricated sheets spontaneously transferring into 3D structures, with programmable bidirectional deformation upon heating and cooling—referred to as pluralized transformation, making NAT-LCEs well-suited for applications in human body-compatible devices and varying environmental conditions. By adjusting the nozzle and plate temperatures, we demonstrate enhanced programmability of NAT-LCEs, exemplified by a wristband designed for improved heart-rate monitoring. This highlights its potential for biomedical and adaptive wearable technologies.

**Results**

To demonstrate the hybrid cooling strategy, we first synthesized a NAT-LCE ink using Michael addition reaction. This process incorporates mixtures of liquid crystal monomers RM82 and RM257, along with the chain extender 2,2'-(ethylenedioxy) diethanethiol (EDDT) (Figure 1B). The resulting ink is comprised of thiol-terminated oligomers designed for crosslinking upon UV exposure, ensuring optimal alignment and curing during the printing process (see further details in Figure S1, Supporting Information). In formulating our LCE inks, we considered factors such as the selection of monomers, their molecular weight differences, and the resulting effects on the nematic-to-isotropic transition temperature ($T_{NI}$) (see further details in Figure S2, Supporting Information). The optimized formulation yielded a $T_{NI}$ of 15.6 °C, ensuring adequate mechanical properties and responsiveness for our applications.

The hybrid cooling strategy involves two critical cooling mechanisms: a liquid-cooled nozzle to control the extrusion temperature ($T_n$) and a cold plate to maintain the substrate temperature ($T_p$). During the printing process, a UV lamp is employed to cure the LCE ink rapidly, locking in the alignment induced by mechanical forces during extrusion and as the ink interacts with the print bed (Figure 1C; see also Figure S3, Supporting Information). The setup effectively solidifies the alignment by carefully controlling UV exposure and printing speed. These combined cooling controls are critical for optimizing the alignment of the liquid crystal mesogens. As shown in Figure 1D, printing at room temperature without cooling results in suboptimal dynamic alignment during extrusion, and the heat generated during UV curing disrupts the static locking of alignment, leading to poorly aligned structures. In contrast, our hybrid cooling-controlled strategy improves dynamic alignments through the cooled nozzle and static alignment locking via the cold plate, resulting in significantly improved alignment. Figure 1F quantifies this improvement, showing a higher-order parameter with our method, confirming its effectiveness in enhancing alignment.

Then, we systematically investigate how shear and extensional forces, when combined with controlled nozzle and cold plate temperatures, enhance the alignment of liquid crystal elastomers during the 3D printing process. By optimizing these temperature conditions alongside the mechanical forces exerted during extrusion, we highlight the crucial role these factors play in both achieving accurate alignment and ensuring the adequate curing of LCEs. Figure 2A illustrates the mechanical forces at play during the extrusion process in 3D printing. As the LCE ink, which contains uncrosslinked liquid crystal oligomers and a photoinitiator, flows through the nozzle, it first experiences both extensional and shear alignment where the ink contacts the nozzle walls, followed by shear alignment inside the nozzle. After being extruded, the ink undergoes extensional alignment as it interacts with the print bed or the previously printed layer. According to the Ericksen-Leslie continuum theory of liquid crystals, these forces cause the liquid crystals' nematic directors to align along the flow direction.[51] The temperature further influences the alignment process, with low-temperature conditions providing a critical advantage. Specifically, lower temperatures slow down the liquid crystal mesogens' relaxation dynamics, thereby preserving the shear-induced alignment more effectively.

This theoretical understanding is quantitatively supported by the data in Figure 2B, which presents the viscosity of the LCE ink across different temperatures (ranging from 5 to 25 °C) under varying shear rates. The viscosity increases as the temperature decreases, indicating that the liquid becomes more flow-resistant. This higher viscosity at lower temperatures is advantageous because it helps maintain the alignment of the liquid crystals induced by the shear forces. The relationship between shear forces and alignment can be described by the following equation for the evolution of the alignment tensor $a_{\alpha\beta}$ under shear:

$$\frac{da_{\alpha\beta}}{dt} = -\frac{1}{\tau}a_{\alpha\beta} + \lambda(\dot{\gamma}_{\alpha\beta} - \omega_{\alpha\beta}) \tag{1}$$

where $\tau$ represents the relaxation time, $\lambda$ is the flow alignment parameter, $\dot{\gamma}_{\alpha\beta}$ is the shear rate tensor, and $\omega_{\alpha\beta}$ is the rotation tensor. The increase in viscosity at lower temperatures ($\tau$ increases) indicates a reduction in the rate at which the system returns to equilibrium, thus favoring the maintenance of the flow-induced alignment.

Figure 2C further corroborates this theory by examining the time required for the viscosity to stabilize under a constant shear rate (10 s$^{-1}$) across the same range of temperatures. The data show that lower temperatures facilitate faster viscosity stabilization, corresponding to quicker and more effective alignment of the liquid crystals. This behavior aligns with the theory that the alignment tensor $a_{\alpha\beta}$ stabilizes more rapidly in cooler environments due to reduced thermal agitation, leading to a more ordered structure in the final printed product. The importance of maintaining low temperatures during the curing process is highlighted in Figure 2D (see also Figure S4, Supporting Information). Here, the thermal effects of UV curing on LCEs placed on a cold plate at 5 °C versus those at room temperature are compared. The images demonstrate that UV curing at room temperature causes a rapid increase in temperature, with the LCE reaching up to 41.7 °C under 30mW/cm$^2$ of UV light, far above the $T_{NI}$ of the LCE ink (15.6 °C). This rise in temperature can disrupt the alignment of the liquid crystals, reversing the beneficial effects of shear-induced alignment achieved during extrusion and extensional alignment. Conversely, when the curing is performed on a cold plate at 5 °C, the low initial temperature reduces the risk of alignment disruption. In contrast, the cold plate efficiently dissipates the heat generated during the reaction, leading to a minimal temperature increase and better alignment preservation. In this case, the cold plate's ability to rapidly dissipate heat and maintain a low temperature effectively solidifies the alignment, preventing polymer chain rearrangement that could occur with slower or insufficient cooling. The curing process is further illustrated by Figure 2E, which shows changes in viscosity over time as the LCE ink is exposed to different intensities of UV light at 5 °C and room temperature. Curing at 5 °C requires higher UV intensities than at 25 °C to compensate for the slower curing speed in cooler conditions. The cold plate environment helps minimize thermal increases associated with higher UV intensities, as shown in Figure 2D. Conversely, increasing UV intensity at room temperature offers limited benefits for alignment due to the heightened risk of misalignment from thermal agitation. The viscosity data confirms that both conditions eventually reach a stable plateau, indicating effective curing within the operational window of 3D printing. In Figure S5 (Supporting Information), we employed a molding technique to create unaligned LCE sheets that were cured for 160 seconds under both 5 °C and 25 °C conditions with UV intensities of 40 mW/cm². The stress-strain curves and Young's modulus at a 35% strain of these cast sheets indicate similar mechanical properties, demonstrating a comparable degree of curing across the two temperature conditions.

To demonstrate the impact of the hybrid cooling controlled strategy on enhancing 3D printing and alignment of NAT-LCEs, LCE bilayer sheets (100 μm thick, 20 mm × 8 mm) were printed under nozzle and plate temperatures of 5 °C and room temperature (25 °C), respectively. The sheets were patterned on a glass substrate using a nozzle diameter of 400 μm and a print speed of 25 mm/min. Different applied pressures were used to ensure the printed filament width remained consistent across all samples. Notably, the LCE samples printed at 5 °C exhibited improved director alignment along the print path, as evidenced by a higher degree of polymer chain alignment. The alignment of the printed LCE samples was characterized using wide-angle x-ray scattering (WAXS) measurements, as shown in Figures 2F and S6, where we present data from five different temperature conditions. Figures 2G and S6 display normalized intensity data plotted as a function of the azimuthal angle, illustrating the degree of polymer chain alignment under these varying conditions. Furthermore, the calculated order parameters for each sample are depicted in Figure 2H. One of the most critical parameters for describing the actuation properties of an LCE is the magnitude of its actuation strain. To assess this, deformation comparisons were made of LCE sheet samples printed under four combinations of nozzle and plate temperatures ($T_n$ and $T_p$) through heating and cooling tests (Figure 2G; see also Figure S7, Supporting Information). The images reveal that samples printed at a lower nozzle temperature of 5 °C and a plate temperature of 5 °C exhibited more significant deformation than those printed under higher temperature conditions. The actuation strain ($\varepsilon_a$) along the printing path is defined as $\varepsilon_a = -(l - l_0)/l_0$ where $l_0$ is the original length of the LCE sheet (at 5 °C), and $l$ is its length at 60 °C. Figure 2H presents a heat map of the actuation strain experienced by LCE sheets under varying printing conditions, highlighting that low-temperature conditions enhance the shear-induced alignment of liquid crystal mesogens during extrusion and preserve this alignment during the subsequent UV curing process. These results confirm that effective control of both $T_n$ and $T_p$ during 3D printing is essential for optimizing the functional properties of LCE materials. To further elucidate the mechanical properties of these printed structures under different cooling temperature conditions, we conducted tensile tests on the aforementioned samples. In these tests, the applied tensile force corresponds to the orientation direction of the polymer chains. The stress-strain curves for these samples are presented in Figure S8, and the corresponding Young's modulus results (at a 35% strain) are shown in Figure 2K, highlighting a clear trend where samples with a higher degree of alignment exhibit increased Young's modulus. Specifically, the samples printed at $T_n$ = 5 °C and $T_p$ = 5 °C demonstrated the highest Young's modulus. In contrast, those printed at $T_n$ = 25 °C and $T_p$ = 25 °C showed the lowest values, confirming the correlation between polymer chain alignment and mechanical properties. The shear field is another key parameter that can impact the LCE alignment. We next conducted our experiments at temperature settings of $T_n$ = 5 °C, $T_p$ = 5 °C. By using various nozzle diameters (1.2 mm, 0.8 mm, and 0.4 mm) and print speeds (5 mm/min, 15 mm/min, and 25 mm/min), we measured the actuation strain, as depicted in Figure S9. Our results indicate that smaller nozzle diameters and faster print speeds enhance the alignment of polymer chains due to greater shear forces, leading to improvements in both actuation strain and Young's modulus along the orientation direction, as further illustrated in Figure S10.

Next, we demonstrated the pluralized yet controllable transformation enabled by hybrid cooling printing. After cross-linking, the liquid crystal elastomers (LCEs) exhibit a noticeable shift in the nematic-isotropic transition temperature ($T_{NI}$), with the $T_{NI}$ of the cured LCE increasing significantly (Figure 3A). Under the influence of UV light during the 3D printing process, photopolymerization is initiated, locking the alignment of the liquid crystal (LC) mesogens that were oriented by shear forces. The structure is designed and printed in a flat configuration, corresponding to a midpoint temperature closely related to the nematic-isotropic transition ($T_{NI}$). As shown in Figure 3B, when the temperature decreases below this midpoint, the liquid crystal mesogens become more ordered compared to their state during printing, resulting in the printed LCE filaments elongating along the alignment direction. Conversely, when the temperature exceeds the midpoint, the liquid crystal mesogens lose their alignment and become disordered. This leads to a different deformation, where the LCE strip shortens, driven by the isotropic phase. This temperature-dependent behavior (enhanced order at lower temperatures and increasing disorder at higher temperatures) explains the reversible shape morphing of the LCE structure. The ability of this method to create sophisticated shapes is attributed to the shifting of the $T_{NI}$ from below to above room temperature during the on-the-fly UV curing in hybrid cooling printing, which enables the NAT-LCE planar structures to transform into 3D shapes spontaneously. This suggests substantial potential for advanced programmable deformation strategies based on variations in the nematic-to-isotropic transition temperature ($T_{NI}$). It is feasible to utilize shifts in the transition temperature as a mechanism to generate elaborate configurations in their as-printed state under ambient conditions. Depending on the programmed printing path, the 2D designs can transform into more intricate 3D shapes, including curved surfaces, thereby streamlining manufacturing processes and enhancing operational efficiency. Moreover, because the deformation behavior spans room temperature, the printed objects exhibit opposite shape changes upon cooling and heating, demonstrating their ability to undergo bidirectional deformation under varying thermal conditions.

Figures 3C-E and Movie S2-4 (Supporting Information) present spontaneous 3D structure formation and bidirectional deformation results. Figure 3C shows an LCE disk with a circular alignment pattern. After printing, the disk spontaneously transforms into a saddle shape (room temperature). It can be calculated from Figure S7 that the strain along the nematic order after release is 14%. According to this, the as-printed 3D shape can be predicted with FEA (Figure S11, Supporting Information). If the disk cooled to 10 °C, the NAT-LCEs further elongate along the nematic order, resulting in increased curvature of the saddle structure. On the other hand, if the temperature increases to 60 °C, it transforms into a cone shape, demonstrating a reversible shape change driven by the programmed nematic order (see also Figure S12, Supporting Information). Figure 3D illustrates a more complex printed structure where the top and bottom layers of the LCE have orthogonal mesogen orientations. At room temperature, the bilayer rectangular LCE sheet twists spontaneously. The orthogonal alignment between the layers induces incompatible strains upon heating or cooling, causing the structure to twist in one direction when heated and in the opposite direction when cooled, effectively showcasing

the programmability of these deformations (see also Figure S13, Supporting Information). Lastly, Figure 3E showcases a grid-patterned LCE structure. At lower temperatures, the grid expands as both the longitudinal and transverse fibers undergo elongation, while at higher temperatures, it contracts (see also Figure S14, Supporting Information). This behavior highlights the LCE's capability for reversible, complex shape changes, which is critical for soft robotics and dynamic systems applications.

By controlling both $T_n$ and $T_p$ during the printing process, the nematic order can be tuned to be graded, enabling the creation of NAT-LCEs with graded properties that facilitate intricate active morphing behaviors. Since both $T_n$ and $T_p$ are negatively correlated with the nematic order of the printed LCEs, these temperatures can be adjusted during the printing of different sections of a single object, resulting in sophisticated deformation patterns. Specifically, we design and print bilayer LCE structures that form the letters "USTC", utilizing a targeted approach to achieve the desired deformation. As illustrated in Figure 4A, the printing process involves programmed temperature control, enabling the creation of these patterned shapes. Each letter is formed through a combination of top and bottom layers with distinct printing parameters, resulting in dynamic morphing behaviors when exposed to thermal stimuli.

Figures 4B to 4D explore different printing strategies to achieve varied morphing behaviors in LCE strips. In Figure 4B, the strips are designed with one bottom layer exhibiting negligible actuation strain and a top layer with maximum actuation strain along the length direction. Upon immersion in 60 °C water, the strips uniformly curl due to the contraction mismatch between the two layers. Figure 4C and Movie S5 (Supporting Information) introduce a graded actuation strain into the top layer by controlling the printing temperature and the existing contraction mismatch between the layers. As a result, each strip bends uniquely when heated, showcasing distinct bending morphologies due to the varying planar gradients of actuation strain within the top layer. The experimental results, shown alongside the FEA simulations in Figure 4D, demonstrate a strong correlation between the predicted and observed deformations. Moreover, Figure 4E illustrates the pluralized transformation in the LCE sheet fabricated by tuning the cooling temperature. The outer ring is printed at room temperature, while the central bar is constructed using a low-temperature nozzle and cold plate, resulting in a higher orientation. After releasing from the printing platform, the poorly oriented outer ring does not undergo deformation, while the central bar spontaneously elongates, causing the central bar to curve up. The bar at low temperatures (10 °C) elongates further, leading to a more pronounced curvature. As the temperature increases, the bar shortens, whereas the outer ring exhibits minimal deformation. The shortening of the bar creates a tensile force, causing the LCE sheet to transform into a saddle shape, exhibiting bidirectional deformation in response to heating and cooling. This demonstrates the potential for creating dynamic, responsive materials that can be engineered for various applications.

Our 3D-printed NAT-LCE holds great promise in bio-compatible devices and robotics. As a demonstration, we developed a wristwatch system with heart rate monitoring and a PID control heating circuit; the heating circuit is connected to a body-compatible, adaptive, and interactive LCE wristband for enhanced heart rate monitoring. Figure 5A presents an exploded view of the wristband, which integrates flexible liquid metal heating elements embedded in silicone (the fabrication method is detailed in Figure S15, while the circuit diagram of the wristwatch system is provided in Figure S16, Supporting

Information) and is controlled by a temperature sensor via a PID control system. This design allows precise regulation of the LCE wristband's contraction and expansion, automatically adjusting its tightness. Figure 5B shows photographs of the wristband with top and side views of it being worn on the wrist. In the relaxed state, the wristband is designed to ensure breathability and comfort during wear, leveraging the strength and flexibility of the LCE grid film. When accurate heart rate monitoring is required, the system can manually or automatically heat the wristband, causing it to tighten around the wrist. Pulling force tests proved that the wristband has sufficient contraction capability (see pulling force results in Figure S17, Supporting Information). Figure 5C displays thermal imaging sequences at different heating intervals (0s, 2s, 10s, 100s), showing the temperature changes over time. As the liquid metal heater warms the wristband to 50 °C, the thermal images reveal a gradual and uniform temperature distribution across the wristband. The heating causes the LCE strip to contract, tightening the wristband. The PID control system modulates the pulse width of current(2.5A), rapidly raising the temperature to 50 °C and stabilizing, ensuring consistent heating during the measurement period (Figure 5D). After the measurement, the system cools down, and the wristband returns to relaxed (see cyclic testing results in Figure S18, Supporting Information).

Photoplethysmography (PPG) signals are collected via a photodiode sensor and utilized in heart rate computation in the MCU[52], and data are also sent to a monitoring device. Users can manually or programmatically control the wristband's tightness by switching on the heater when more accurate data is required, causing the wristband to tighten (Figure 5E). This tightening significantly improves the accuracy of the heart rate measurement by reducing noise compared to the loose state. Figures 5F and 5G and Movie S6 (Supporting Information) compare the heart rate measurement performance in loose and tight conditions. Figure 5F shows data from the system without heating, where the wristband remains loose, leading to increased noise and lower accuracy. In contrast, Figure 5G shows the amplitude and heart rate data when the wristband is heated to 50 °C and tightened around the wrist, resulting in improved measurement precision and reduced noise. The comparison of amplitude and heart rate data between the two states demonstrates the significant enhancement in signal quality when the wristband tightens, validating the system's ability to adjust tightness to improve heart rate monitoring performance dynamically. In Figure 5H, we demonstrate the scalability of the wristband by showcasing data collected from an individual performing various activities, such as working, dining, walking, and snapping. The data illustrates the performance variability of the wristband in both relaxed and tightened states (the first 20 seconds showcase relaxed states, while the subsequent 20 seconds correspond to tightened states). We observed a significant reduction in noise and an improvement in data quality during these activities. Additionally, we conducted a 1000-cycle fatigue test, the results of which are included in Figure S19 in the Supporting Information. Remarkably, after 1000 cycles, the tensile performance of the wristband demonstrated no significant deterioration, highlighting its robustness for practical applications.

**Conclusion**

This study demonstrated a hybrid cooling strategy to programmable 3D printing of near-ambient temperature-responsive liquid crystal elastomers (NAT-LCEs) with enhanced nematic order, intricate shape forming, and morphing capability. By incorporating both a liquid-cooled nozzle and a cold substrate plate, we effectively improved the alignment of liquid crystal mesogens during printing, ensuring optimal curing and minimizing thermal disruptions. This approach achieved high nematic order and facilitated the creation of complex, programmable structures capable of bidirectional deformation at near-ambient temperatures. Our findings reveal the potential of this hybrid cooling strategy in advancing the application of NAT-LCEs in various fields, such as soft robotics, biomedical devices, and adaptive wearable technologies. Future work will focus on refining the printing parameters and exploring the integration of NAT-LCEs with other innovative materials to further expand their functional capabilities.

**Experimental Section/Methods**

*Ink Preparation*: The LCE was synthesized using a previously reported Michael addition method.[29,53] As-received 2,2′-(ethylenedioxy) diethanethiol (EDDT, Sigma-Aldrich), RM82, 1,4-bis-[4-(3-acryloyloxypropyloxy) benzoyloxy]-2-methylbenzene (RM257, Zhende Chemical Technology Inc.), and 1,3,5-triallyl-1,3,5-triazine-2,4,6(1H,3H,5H)-trione (TATATO, Sigma-Aldrich) was added into a 25 mL round-bottom flask in a mole ratio of 1.0:0.4:0.4:0.133, followed by adding 1 wt% triethylamine (TCI), 2 wt% butylated hydroxytoluene (Sigma-Aldrich), and 1.5 wt% Irgacure 651 (BASF). The molar ratio used was 0.8 acrylate:1.0 thiol:0.2 vinyl. All materials were melted with a heat gun and blended for 5 min. Then, the flask was filled with nitrogen and stirred to complete prepolymerization for 3 h at 65 °C without light, forming LC ink. The LCE with $T_{NI}$ = 15.6 °C was stored at -18 °C before printing.

*Rheological characterization*: Rheological properties of NAT-LCE inks were measured by a rotational rheometer (Physical MCR 302, Anton Paar, Austrian), and a 25 mm steel Peltier plate was used with a 0.5 mm gap distance. For each test, samples were heated to 120 °C to erase the temperature history and kept at the measurement temperature (i.e., 5 °C/10 °C/15 °C/20 °C/25 °C) for 5 minutes before measurement. The shear rate was swept from $10^{-2}$ s$^{-1}$ to $10^{2}$ s$^{-1}$ during the viscosity measurement. The viscosity change under UV explosion was conducted with a 0.1 mm gap distance. A glass base of the rheometer was used to allow UV light (generated by Omnicure S2000, Excelitas) to pass through, with a light intensity ranging from 40 to 60 mW/cm² at 5 °C and from 20 to 40mW/cm² at 25 °C. The UV intensity was measured using an optical power meter (S120VC, THORLABS, USA). During the measurement, a constant shear rate of 0.1 s$^{-1}$ was applied.

*Differential scanning calorimetry (DSC) tests*: DSC tests were conducted using TA Instruments (Q2000 DSC, USA). Approximately 5 mg of each sample was sealed in aluminum pans and analyzed in a nitrogen atmosphere over a temperature range of -50 °C to 150 °C, with a heating rate of 10 °C/min. The analysis involved a heat-cool-heat cycle to remove thermal history during the first heating ramp and assess the glass transition temperature ($T_g$) and the inks' nematic-to-isotropic transition temperature ($T_{NI}$). Samples were held isothermally for 1 minute at both high and low temperatures. Data from the second heating ramp determined the $T_g$ and $T_{NI}$ values.

*Hybrid Cooling 3D Printing for NAT-LCEs:* The hybrid cooling 3D printing of NAT-LCEs was performed using a modified commercial fused deposition modeling (FDM) printer (Ultimaker 2+ Extended, Ultimaker, Netherlands). This printer employs a pressure-driven extrusion system controlled by a digital pneumatic regulator (SuperΣ CMIII, MUSASHI Engineering, Japan) to deposit ink based on programmed G-code instructions. The printing paths were derived from custom-designed STL models created in SolidWorks (Dassault Systèmes, France) and converted into G-code using the commercial software Ultimaker Cura (Ultimaker, Netherlands). Precleaned glass plates are used as a printing platform. Stainless steel nozzles with an inner diameter of 400 μm were used for all printing experiments. A custom-fabricated ring-shaped water cooling pipe made of aluminum alloy was 3D printed, and a liquid chiller was manufactured using selective laser sintering (SLS) technology to conform to the shape of the nozzle. An embedded K-type thermocouple is mounted on the printhead, and the nozzle temperature is set to the print temperature and maintained with a temperature controller. A thermal and light-insulating cover is placed over the cooling head to minimize heat dissipation and prevent the LCE ink within the barrel from curing due to exposure to UV light (Figure S3, Supporting Information).

Before printing, all prepared inks were transferred into 10 cc barrels (model PSY-10E, manufactured by MUSASHI Engineering, Japan). The inks were then heated above the isotropic transition temperature ($T_{NI}$) and centrifuged at 2200 rpm five times to remove any air bubbles. The system was maintained at the designated printing temperature (i.e., 5 to 25 °C) for approximately 5 minutes before printing to achieve a steady-state operating condition. During extrusion, the ink was exposed to UV light (Omnicure S2000, Excelitas) at an intensity of about 40 mW/cm². After printing, the LCE was further post-cross-linked under UV exposure (≈40 mW/cm²) for 15 minutes on each side to ensure uniform crosslinking.

*Characterization of printed LCEs:* Alignment of printed LCE samples was characterized by x-ray scattering measurements on a SAXSLAB system with a Rigaku 002 microfocus x-ray source ($\lambda = 1.5409$Å) with sample to detector (PILATUS 300K, Dectris) distance of 83.3 mm for 20 min to capture the mesogens-mesogen correlations at $q\sim1.5$Å$^{-1}$. Wide-angle x-ray scattering samples comprised two-layer printed LCE unidirectional strips (180 to 200 μm thick). The nematic orientation parameter was characterized by Hermans' orientation parameter (S), given by equations 2 and 3,

$$\langle cos^2\phi \rangle = \frac{\int_0^\pi I(\phi)cos^2\phi \sin\phi \, d\phi}{\int_0^\pi I(\phi)\sin\phi \, d\phi} \qquad (2)$$

$$S = \frac{3\langle cos^2\phi \rangle - 1}{2} \qquad (3)$$

where $\langle cos^2\phi \rangle$ is the average cosine square of the angles between the long axis of individual mesogens and the global LC director. $I(\phi)$ is the angle-dependent scattering intensity from the WAXS patterns.[54] Orientation parameter calculations were performed using Matlab (Mathworks, Natick, MA).

*Mechanical properties.* The specimens for mechanical tests were printed or cast into a dogbone specimen with a cross-section of 0.4 mm × 4 mm. Tensile properties were measured on a universal testing machine with a constant crosshead speed of 30 mm/s$^{-1}$ (AGS-X, SHIMADZU, Japan). Additionally, tensile tests were conducted on five replicates of each sample condition to ensure statistical reliability.

To measure the contraction strain of LCE sheets, the samples were placed on a stainless steel temperature-controlled plate covered with silicone oil (PMX-200, Dow Corning) to prevent adhesion to the substrate and enhance heat transfer before actuation. The original length of the sample (denoted as $l_0$) was measured and compared with the contracted length (denoted as $l$) at different temperatures, which were captured using an infrared camera (ETS320, FLIR, USA) (Figure S7, Supporting Information). The heat-induced contraction strain was calculated based on these measurements. The temperature history was tracked using FLIR Tools+ software, as the sample was heated from 5 to 70 °C and cooled down to 5 °C, with a heating/cooling rate fixed at 2 °C/min for all tests. Optical images were captured using a CMOS camera (acA2440-20 gm/gc, Basler, Germany), and geometrical information was measured using ImageJ software. For each sample, three independent measurements were conducted.

For the characterization of deformation and demonstration experiments, the thermally induced actuation of printed LCE structures, as shown in Figure 4, was conducted in a glass Petri dish filled with warm water (~60 °C).

*FEA of the deformation of 3D-printed LCE structures:* FEA is conducted using commercial software Abaqus. LCE is modeled as a linear thermoelastic material with an anisotropic thermal expansion coefficient and Poisson's ratio of 0.499. The thermal expansion coefficient (is negative) in the direction parallel to the axial direction of the printed LCE filament is obtained from the measurements shown in Figures 3 and 4.

*Fabrication of the wristwatch.* The wristwatch system consists of a watch screen, battery, circuit boards, and an LCE wristband that incorporates flexible liquid metal heating elements embedded in silicone. This design enables controlled heating, allowing for the contraction and expansion of the LCE wristband. The watch casing is fabricated from FDM 3D-printed materials. Using high-precision 3D printing, a mold is created for casting Eco-flex 30. This process produces an S-shaped flexible heating wire with embedded microchannels into which EGaIn liquid metal is injected. The liquid metal acts as a conductive heating material. It is encapsulated with a PID control heating circuit and temperature sensor, ensuring that it does not interfere with the stretching and contraction of the LCE strip. The integration of the heating element, temperature sensor, and LCE strip is achieved using moisture-curing silicone Rubber (ELASTOSIL E41).

*Workflow of the controller.* The wristwatch's control system consists of a microcontroller unit (MCU), a photoplethysmography (PPG) acquisition module, a Bluetooth module, and a heating circuit. The MCU is responsible for processing PPG signals from the acquisition module, computing heart rate, and transmitting the data to a computer via Bluetooth. It also generates pulse width modulation (PWM) signals to control the heating circuit through a CMOS switch. PWM controls the heating speed by varying the duty cycle, which is the proportion of high-voltage pulse in a periodic signal. In this experiment, the temperature is fed back to the MCU via a thermistor; an analog-to-digit converter (ADC) contained in the MCU senses the voltage division across the thermistor to calculate its current resistance and the corresponding temperature. The frequency of the PWM signal is set at 5Hz. When the temperature is below 45 °C, the duty cycle is fixed to 100%, allowing the wristband to heat at full speed. When the temperature exceeds 45 °C, the duty cycle is modulated through a PID system to stabilize the wristband temperature around 50 °C. The PPG acquisition module records the intensity signals of reflected red and infrared light, and artifacts caused by movement are suppressed using template subtraction of the two signals. The heart rate is calculated by detecting the mean peak interval in the continuous signal for 10 seconds.

*Pulling force test methods for the wristband.* The pulling force tests for the wristband were performed using a specialized setup designed to evaluate the performance of the LCE strip. One end of the LCE strip was attached to a load cell (AT8301, AUTODA, China), while the opposite end, along with the load cell, was fixed in place. This configuration enabled the controlled application of tensile forces to the LCE strip. The tests were conducted across a temperature range from 20 °C to 80 °C to assess the influence of temperature on the pulling force. The temperatures during the tests were recorded using an infrared camera (ETS320, FLIR, USA). Additionally, pulling force tests were conducted on five replicates to ensure statistical reliability.

*Fatigue test methods for the wristband.* The fatigue testing setup for the wristband was identical to that used in the pulling force tests. During this analysis, the wristband underwent 1000 cycles of loading and unloading. Each cycle involved applying a current of 2.5 A for 30 seconds, followed by a 60-second cooling period, with this process being controlled by a relay. This cyclic loading approach allowed us to evaluate the material's fatigue resistance under repeated stress effectively.

**Supporting Information**

Document S1. Figures S1-S19, Table S1.

Video S1. Actuation strain in LCE samples printed at different nozzle and substrate temperatures, related to Figure 2

Video S2. Actuation of NAT-LCE printed with a circular alignment pattern, related to Figure 3

Video S3. Twisting deformation of an Actuation of NAT-LCE with orthogonal layers, related to Figure 3

Video S4. Actuation of Grid-patterned NAT-LCE, related to Figure 3

Video S5. Programmable temperature-controlled 3D printing of bilayer LCE structures shaping the letters "USTC", related to Figure 4

Video S6. Body-compatible, adaptive, and interactive LCE wristband for enhanced heart rate monitoring, related to Figure 5

**Acknowledgements**

D.L. and Y.S. contributed equally to this work. This work was supported by the National Key Research and Development Program of China (Grant No. 2020YFA0710100), Natural Science Foundation of Anhui Province (Grant No. 2108085ME170), The Joint Funds from Hefei National Synchrotron Radiation Laboratory (Grant No. KY2090000068), and USTC Research Funds of the Double First-Class Initiative (Grant No. YD2090003002). This work is partially carried out at the USTC Center for Micro- and Nanoscale Research and Fabrication, Instruments Center for Physical Science, University of Science and Technology of China.

**Author Contributions**

Conceptualization, D.L., Y.S., and M.L.; Methodology, D.L., and Y.S.; Validation, D.L., X.L., and Z.Z.; Investigation, D.L., X.L., S.N., and J.W.; Resources, X.L., B.S., and Z.Z.; Writing - Original Draft, D.L.; Writing - Review & Editing, Y.S., W.L., T.P., and M.L.; Supervision, Y.S., S.Z., and M.L.

**Conflict of Interest**

The authors declare no conflict of interest.

# Figures

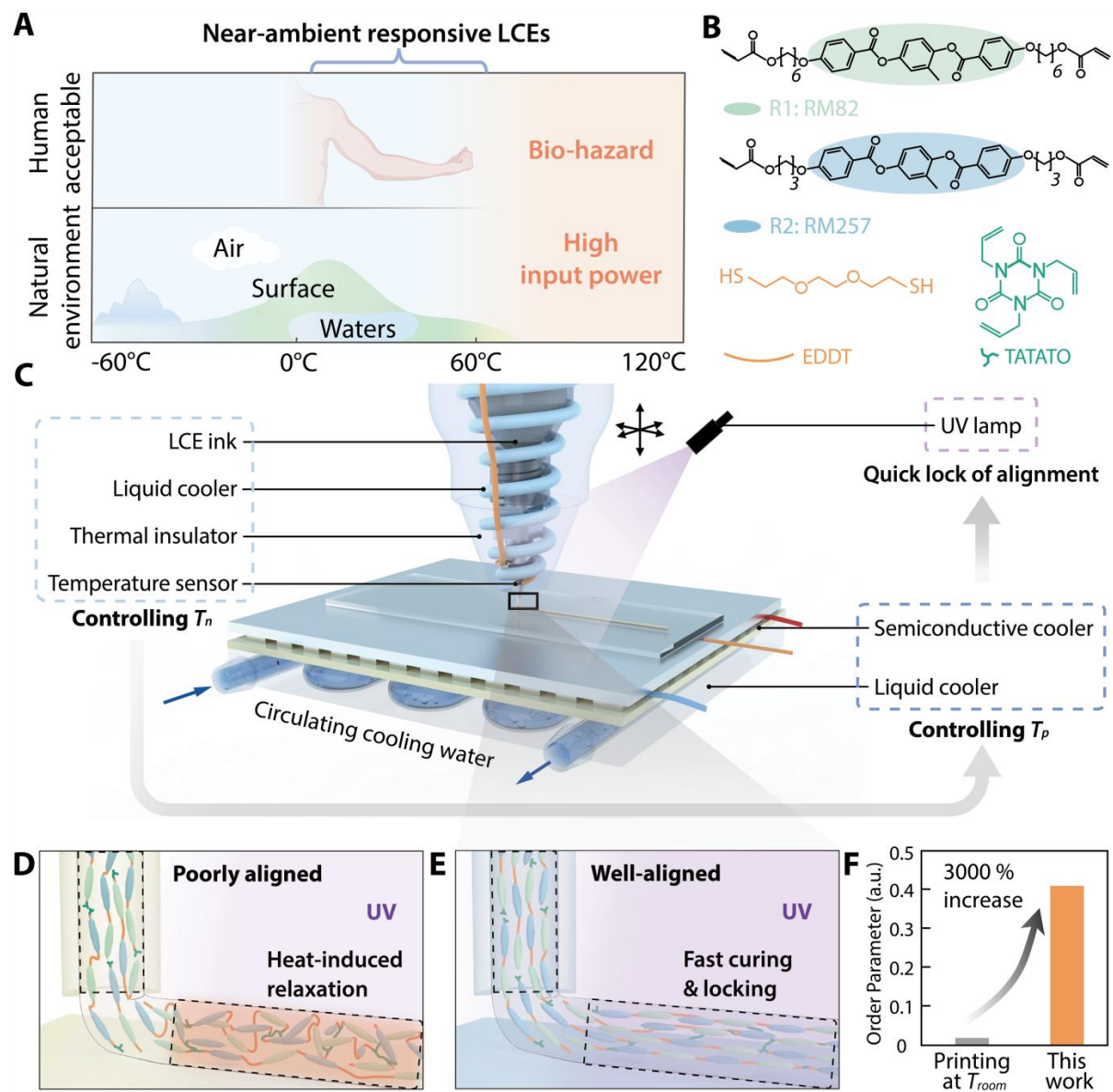

**Figure 1.** Hybrid cooling-controlled strategy for enhanced 3D printing and alignment of near-ambient temperature-responsive liquid crystal elastomers (NAT-LCEs). (A) Temperature ranges for NAT-LCE applications, highlighting their use in near-ambient environments and human-compatible biomedical devices. (B) Chemical structure of mesogenic monomers (RM82, RM257), flexible linker (EDDT), and cross-linker (TATATO) used in the synthesis of NAT-LCE ink. (C) The hybrid cooling system schematic combines a liquid-cooled nozzle and a semiconductive cooler for the print bed. This setup allows controlled alignment of liquid crystal mesogens during extrusion and rapid UV-induced alignment locking. (D) Poor alignment observed in NAT-LCE structures printed without cooling, where UV-induced heat causes relaxation and misalignment of the liquid crystal mesogens. (E) Well-aligned NAT-LCE structures produced using the hybrid cooling strategy allow rapid curing and effective alignment locking. (F) Quantification of the order parameter, showing significant improvement in alignment for NAT-LCEs printed with the hybrid cooling strategy compared to room temperature printing.

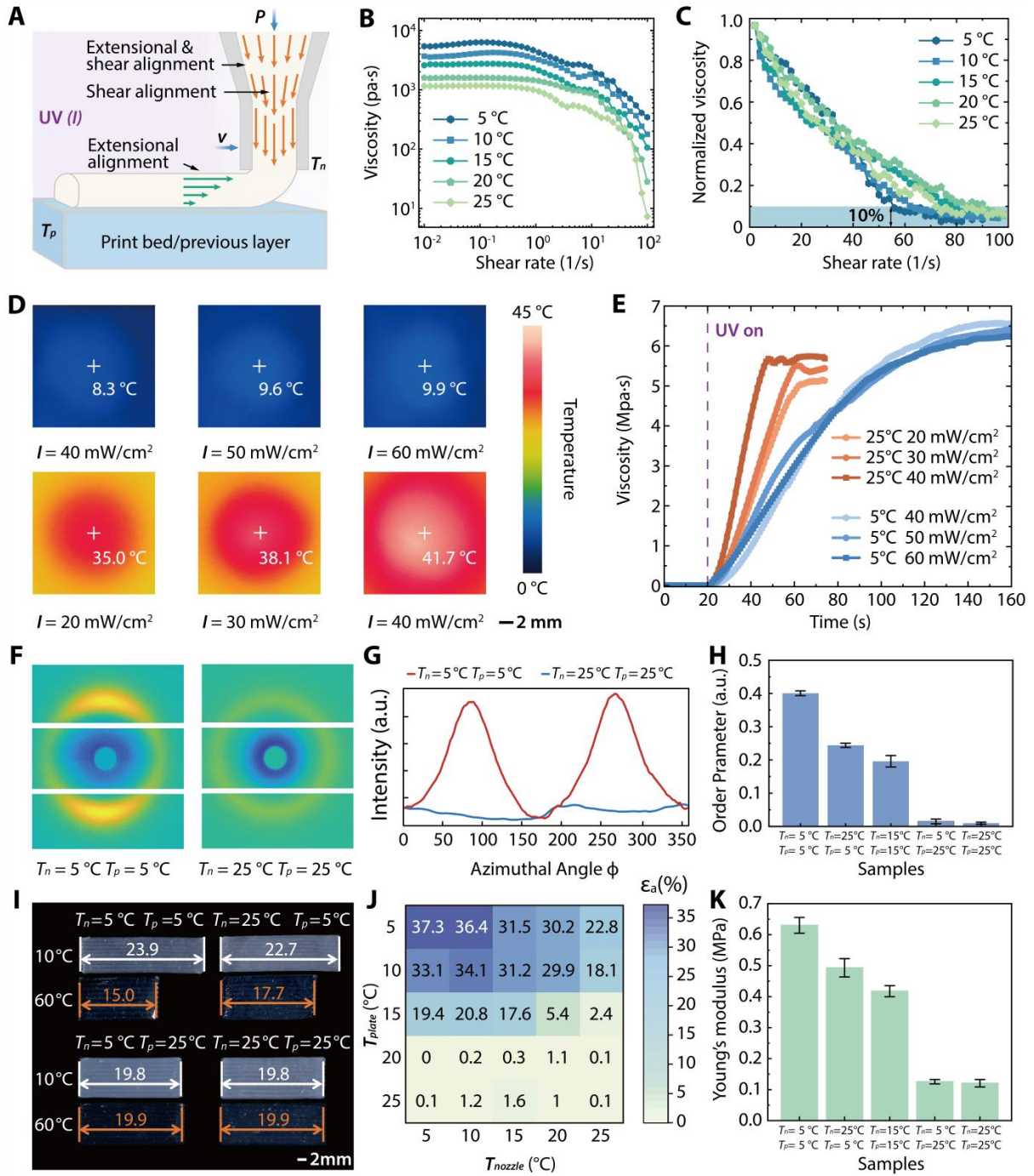

**Figure 2.** Dependence of NAT-LCE alignment on the nozzle and cold plate temperature. (A) Schematic illustration of the shear and extensional alignment forces acting on the NAT-LCE during extrusion and interaction with the print bed. (B) Viscosity of the NAT-LCE ink measured at different temperatures (5 to 25 °C) under various shear rates. (C) Normalized viscosity as a function of shear rate at different temperatures, indicating faster stabilization at lower temperatures. (D) Thermal images showing the temperature rise during UV curing at different light intensities (20 to 60 mW/cm²) for NAT-LCE ink on a cold plate at 5 °C versus room temperature. (E) Viscosity evolution of the LCE ink during UV exposure at various light intensities, ranging from 20 to 40 mW/cm² at 25 °C and from 40 to 60 mW/cm² at 5 °C. The UV exposure begins at 20 seconds. (F) 2D wide-angle X-ray scattering patterns of LCE samples manufactured under different temperature conditions ($T_p$ = 5 °C, $T_n$ = 5 °C and $T_p$ = 25 °C, $T_n$ = 25 °C). (G) Normalized intensity is plotted as a function of the azimuthal angle of LCE samples manufactured under different temperature conditions ($T_n$ = 5 °C, $T_p$ = 5 °C and $T_n$ = 25 °C, $T_p$ = 25 °C). (H) Order parameter of LCE samples manufactured under different temperature conditions ($T_n$ = 5 °C, $T_p$ = 5 °C; $T_n$ = 25 °C, $T_p$ = 5 °C; $T_n$ = 15 °C, $T_p$ = 15 °C; $T_n$ = 5 °C, $T_p$ = 25 °C; $T_n$ = 25 °C, $T_p$ = 25 °C). (I) Comparison of actuation strain in LCE samples printed at different nozzle and substrate temperatures, revealing more significant deformation in samples printed at lower temperatures. (J) Heat map of the actuation strain experienced by LCE sheets under varying printing conditions, showing that low-temperature conditions enhance actuation strain due to better alignment of the liquid crystal mesogens during extrusion. (K) Young's modulus of printed structures under different cooling conditions.

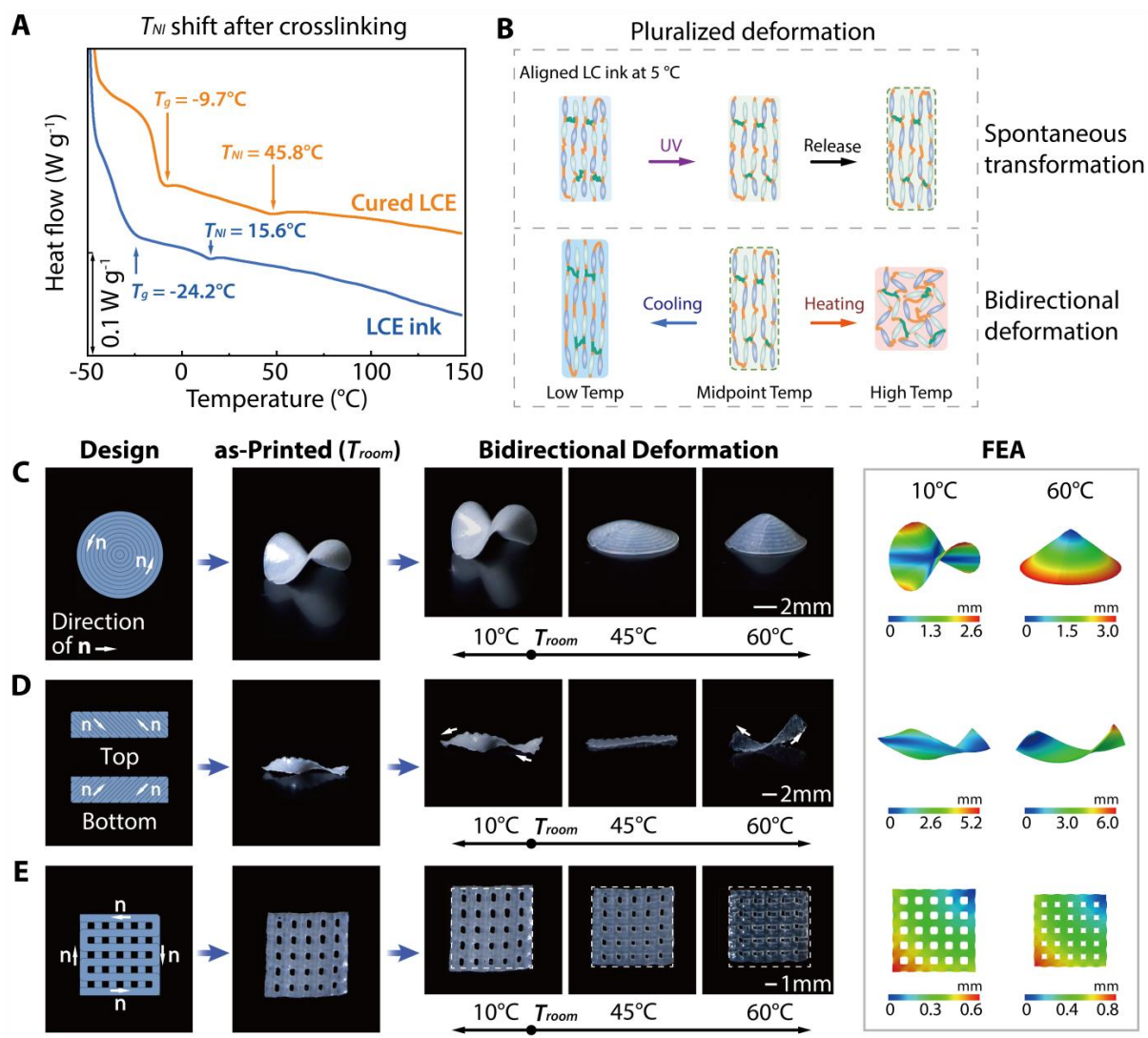

**Figure 3.** Programmable pluralized deformation of NAT-LCE. (A) Differential scanning calorimetry (DSC) showing the shift in the nematic-isotropic transition temperature ($T_{NI}$) after cross-linking. (B) Schematic of the spontaneous transformation and bidirectional deformation of NAT-LCE. At lower temperatures, the liquid crystal mesogens become more ordered, causing elongation along the alignment direction, while at high temperatures, it shortens as the alignment is lost. (C) LCE fiber printed with a circular alignment pattern. It elongates into a saddle shape at low temperatures and transforms into a cone shape at high temperatures. FEA simulations confirm this behavior. (D) Twisting deformation of an LCE structure with orthogonal layers. The structure twists in opposite directions at low and high temperatures, as shown in experiments and FEA simulations. (E) Grid-patterned LCE. Experiments and FEA simulations confirm that the grid expands at low temperatures and contracts at high temperatures.

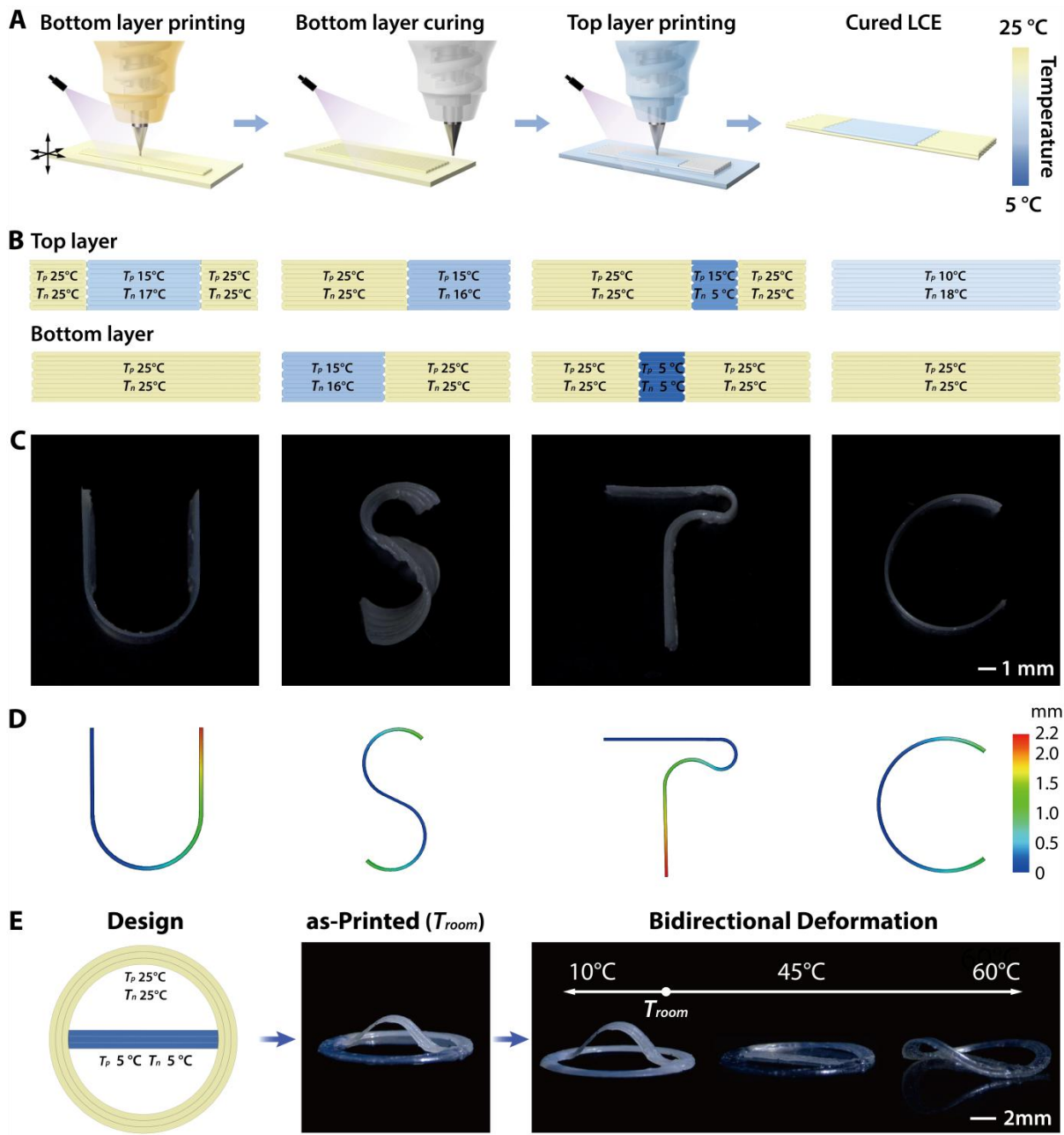

**Figure 4.** Programmable temperature-controlled 3D printing of graded properties NAT-LCE structures. (A) Schematic illustration of the bilayer 3D printing process. The bottom layer is printed and cured, followed by the printing of the top layer, both with distinct temperature-controlled settings. (B) Temperature-controlled printing parameters for the top and bottom layers. (C) Optical images of the printed LCE letters "USTC" after thermal activation, demonstrating the dynamic morphing behaviors at 60 °C due to the bilayer design. (D) Finite element analysis (FEA) simulations showing the deformation patterns of the letters "USTC." (E) Programmable shape morphing LCE sheet spontaneously turning into a 3D structure after release from the platform, exhibiting bidirectional deformation with heating and cooling.

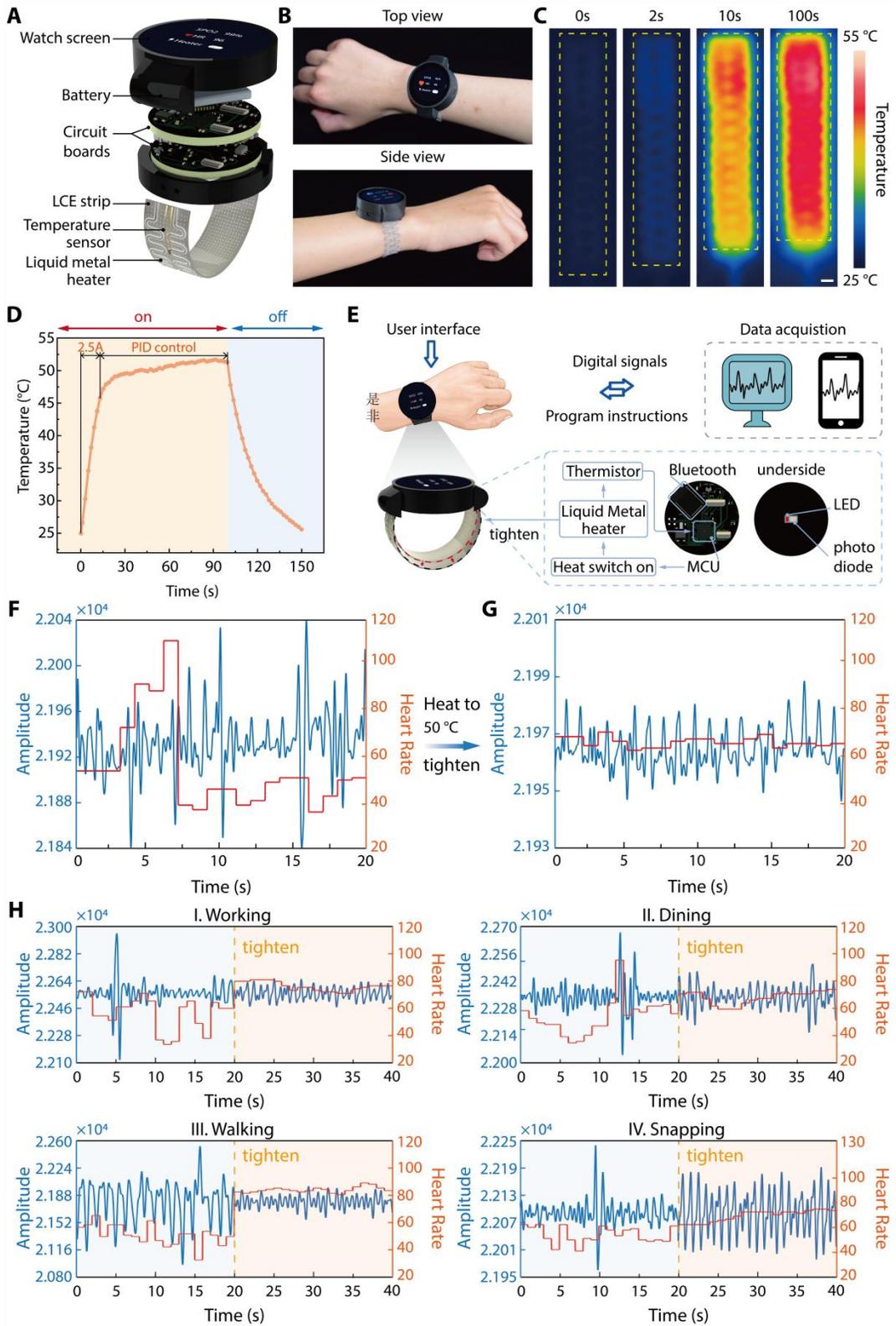

**Figure 5.** Body-compatible, adaptive, and interactive LCE wristband for enhanced heart rate monitoring. (A) Exploded view of the LCE wristband. (B) Photographs of the wristband in top and side views. (C) Thermal images showing the temperature distribution across the wristband at different time intervals (0s, 2s, 10s, 100s) during heating (scale bar = 2 mm). (D) Temperature control graph, showing the wristband reaching 50 °C within 15 seconds under PID control with a 2.5A current. The system stabilizes and cools down after the measurement. (E) Schematic of the data acquisition system, illustrating how the wristband collects heart rate data via sensors and sends signals to a monitoring device through Bluetooth for real-time adjustments. (F) Heart rate monitoring data shows increased noise and lower accuracy when the wristband is loose. (G) Heart rate monitoring data after the wristband tightens (heated to 50 °C) shows improved signal quality and reduced noise. (H) Data was collected from an individual performing various activities (working, dining, walking, and snapping). The amplitude (blue line) and heart rate (red line) are plotted against time(s). The first 20 seconds reflect the relaxed state, while the subsequent 20 seconds indicate the tightened state.